# Human-Centered Artificial Social Intelligence (HC-ASI)

Hanxi Pan[1], Wei Xu[1], Mowei Shen[1], Zaifeng Gao[1]

[1] Department of Psychology and Behavioral Sciences, Zhejiang University, Hangzhou, China

## Abstract

As artificial intelligence systems become increasingly integrated into human social contexts, Artificial Social Intelligence (ASI) has emerged as a critical capability that enables AI to perceive, understand, and engage meaningfully in complex human social interactions. This chapter introduces a comprehensive framework for Human-Centered Artificial Social Intelligence (HC-ASI), built upon the Technology-Human Factors-Ethics (THE) Triangle, which systematically addresses both technical foundations and human-centered design principles necessary for developing socially intelligent AI systems. This chapter provides a comprehensive overview of current ASI research. This chapter begins by establishing the theoretical foundations of ASI, tracing its evolution from classical psychological theories of human social intelligence to contemporary computational models, then examines the mechanisms underlying human-AI social interaction with particular emphasis on establishing shared social understanding and appropriate role positioning. The chapter further explores ASI's practical implications for individuals and groups through comprehensive evaluation frameworks that combine technical benchmarks with human-centered experiential assessments, demonstrating real-world applications through detailed case studies spanning healthcare, companionship, education, and customer service domains. Building on the overview and the framework of HC -ASI, this chapter articulates core HC-ASI design principles and translates them into actionable methodologies and implementation guidelines that provide practical guidance for researchers and practitioners. This chapter concludes with a critical discussion of current challenges and promising directions for developing comprehensive HC-ASI ecosystems.

# 1. Introduction

## 1.1 Overview

The fundamental challenge facing artificial intelligence today is not computational power or algorithmic sophistication, but rather the ability to navigate the subtle complexities of human social interaction. While AI systems excel at pattern recognition, optimization, and knowledge synthesis, they often falter when required to understand social context, interpret emotional nuance, or engage in meaningful interpersonal exchange. This gap between technical capability and social competence has become increasingly apparent as AI systems transition from isolated task performers to integrated social participants in education, healthcare, and collaborative work environments.

Social intelligence, originally conceptualized as a psychological construct, describes the capacity to understand and navigate social relationships effectively. Thorndike (1920) initially defined it as "the ability to understand and manage men and women, boys and girls—to act wisely in human relations" (Thorndike, 1920). Subsequent developments in psychology and cognitive science have expanded this concept into a multidimensional framework encompassing cognitive and affective abilities that enable individuals to maintain relationships, coordinate collective actions, create shared meaning within communities, and navigate complex social environments (Baron-Cohen et al., 1985; Gardner, 1983; Goleman, 2006; Guilford, 1967).

For AI to realize its full potential and integrate harmoniously within human society, it needs to develop analogous social capabilities. This imperative has catalyzed the emergence of artificial social intelligence (ASI)—a field dedicated to endowing AI systems with the capacity to understand, adapt to, and meaningfully participate in human social interactions (Fan et al., 2022; Natu et al., 2023; C. S. Song et al., 2024; Williams et al., 2022). Recent advances in machine learning, natural language processing, and multimodal perception have driven significant progress in ASI development, with contemporary systems demonstrating increasing proficiency in recognizing social cues, modeling social inferences, and generating contextually appropriate responses.

However, as AI systems become embedded in socially sensitive domains—including education, mental health support, and collaborative work environments—new challenges have emerged. These include the computational modeling of nuanced social dynamics, integration of diverse cultural contexts and ethical considerations, ensuring system transparency and user privacy, and mitigating potential unintended social consequences. This chapter provides a comprehensive examination of social intelligence in AI from a human-centered perspective, exploring core concepts, current research progress, key challenges, and societal impacts.

## 1.2 A Framework of Human-Centered Artificial Social Intelligence (HC-ASI)

Given that ASI systems are designed to understand, interpret, and participate in human social contexts, their development should align closely with human-centered AI (HCAI) principles. HCAI emphasizes placing human needs, values, and well-being at the center of system design and development processes. Building upon the Technology-Human Factors-Ethics (THE) Triangle framework proposed by Xu (2019), a Human-Centered AI (HCAI) framework that integrates technological innovation, human factors principles, and ethical considerations, this section presents a comprehensive framework for human-centered ASI (HC-ASI), as illustrated in Figure 1.

The THE Triangle framework addresses three critical dimensions: ethically aligned design, technology enhancement, and human factors design(W. Xu, 2019a, 2019b; W. Xu et al., 2025), providing essential guidance for ensuring that socially intelligent AI systems genuinely serve human needs while operating within appropriate ethical boundaries. The integration of technology, ethics, and human factors should be considered as a coherent sociotechnical system. Each dimension contributes unique but interdependent priorities that together ensure ASI systems remain beneficial, trustworthy, and aligned with human values.

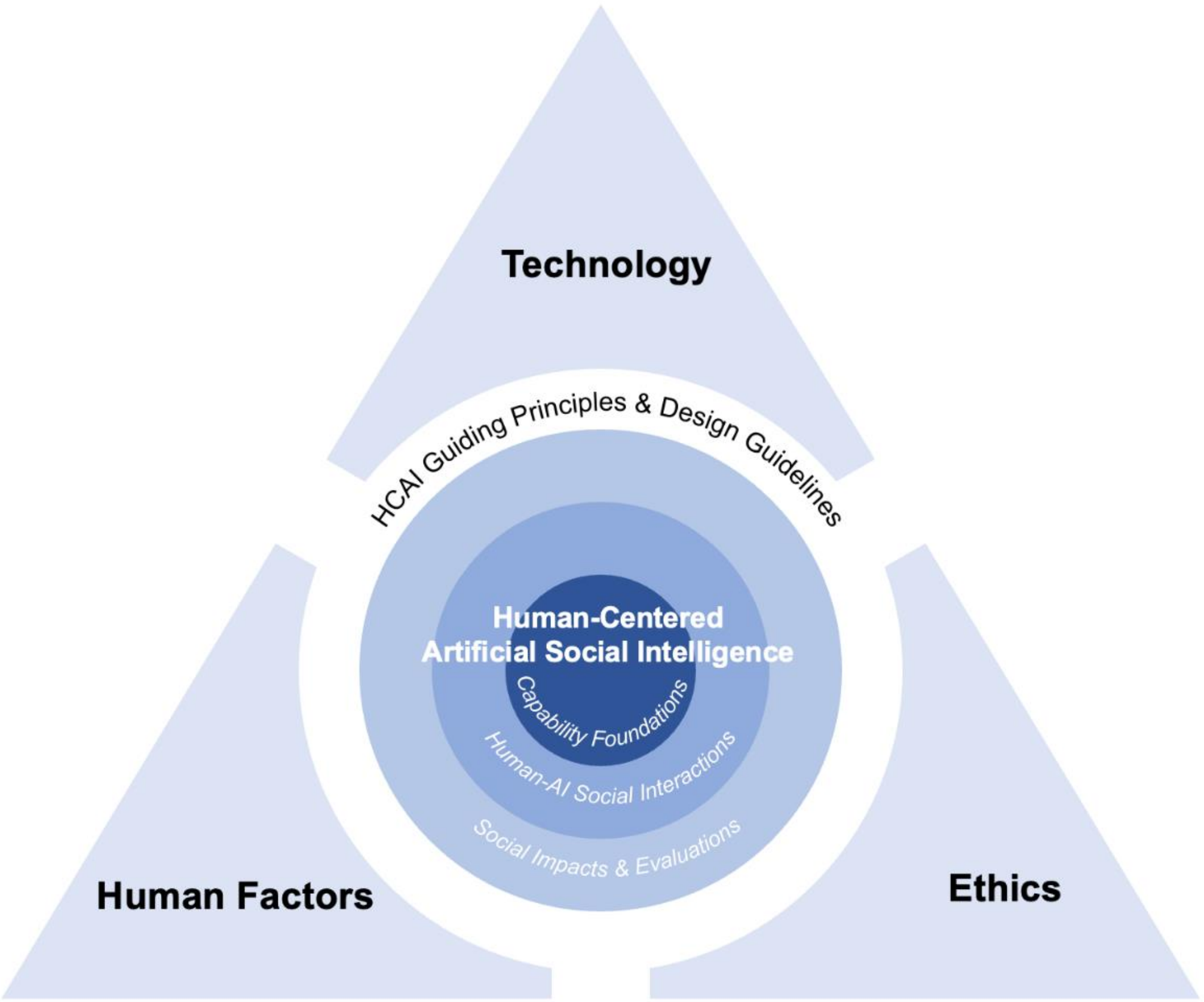

Figure 1 The Framework of Human-Centered Artificial Social Intelligence (HC-ASI)

### *1.2.1 Technology Perspective*

The technology perspective of THE framework emphasizes augmenting human capabilities through the synergistic combination of human and machine intelligence. ASI development should draw foundational inspiration from human social intelligence while achieving deep integration with human social environments. This enables AI systems to function effectively in complex, dynamic social scenarios and serve as genuine social partners that enhance human capabilities.

The theoretical design of ASI should be grounded in theories of human social intelligence. While purely data-driven approaches demonstrate excellence in pattern recognition, they exhibit limitations when addressing the complexity and context-dependency inherent in social interactions. The hybrid intelligence framework proposed by Akata et al. (2020) emphasizes that the design of AI social capabilities should draw upon the profound understanding of human social cognitive mechanisms developed within psychology, sociology, and cognitive science. By integrating these theories and incorporating human experts' social knowledge, cultural norms, and ethical principles, ASI can more accurately interpret social cues, infer intentions, generate norm-compliant behaviors, and navigate complex cross-cultural situations such as politeness conventions and social customs.

Hybrid human-AI intelligence extends beyond AI imitation of human social intelligence; it aims to design systems that complement human strengths. AI should enhance human cognitive capacities in social interactions—for instance, by analyzing conversational patterns to help users identify social cues or providing context-aware support during group collaboration. In practice, ASI systems need to possess deep understanding of human social behavior while maintaining a supportive role toward human judgment and decision-making, prioritizing human goals and social development.

To effectively integrate human-AI intelligence, HC-ASI need to engage through bidirectional interaction within complex and dynamic human social environments. The "socially situated artificial intelligence" proposed by Krishna et al. (2022a) emphasizes that AI should learn social behaviors from authentic human interactions through feedback mechanisms, model updating strategies, and knowledge accumulation methods. For instance, in team collaboration, social AI should monitor dynamics, identify potential conflicts, and provide mediation suggestions while maintaining human agency.

### *1.2.2 Human Factors Perspective*

This perspective requires that AI systems be explainable, comprehensible, useful, and usable. HC-ASI systems should demonstrate transparent and reasonable behavior that aligns with human social logic, establish genuine shared understanding with users, and undergo human-centered evaluation that captures both cognitive and affective outcomes.

Given the nuanced nature of social interaction, HC-ASI systems need to exhibit transparent and reasonable behavior aligned with human social logic. This means AI social behavior should not merely be technically correct but socially sensible, conforming to human intuitive understanding and expectations of social interaction. Consistency with social logic manifests at multiple levels: AI behavior should follow social norms and etiquette conventions, such as using appropriate politeness markers, maintaining proper personal space, and expressing empathy at suitable moments. Moreover, AI decisions should possess social intelligibility—when AI makes judgments or takes actions in social contexts, the underlying reasoning should be explainable through social rationale rather than mere algorithmic outputs.

Establishing shared understanding serves as the foundation for effective social interaction and the key pathway to achieving system explainability and usability (Baek & Parkinson, 2022; Reis et al., 2017). For social AI, shared understanding of social contexts holds particular importance, as social information inherently contains ambiguity and polysemy—identical social cues may convey entirely different meanings across contexts(Azaad & Sebanz, 2023; Thornton, 2024). Therefore, humans and social AI should reach consensus on the perception, inference, and response to social situations, ensuring AI behavior appears predictable, comprehensible, and socially logical to users.

The human factors perspective further emphasizes that ASI cannot be evaluated solely by technical metrics but should also focus on user subjective experience and actual social impact. ASI evaluation should center on its unique performance in social interaction: how AI demonstrates social intelligence in authentic social contexts, how it affects interpersonal interaction quality, and how users perceive and accept it at the social level. Particularly crucial is its facilitation of meaningful human-AI social interaction and the impact on social relationships, such as AI's influence on users' social skill development, interpersonal relationship quality, and psychological well-being. These comprehensive evaluation approaches ensure ASI design genuinely centers on user needs and experiences rather than merely pursuing technical performance improvements.

### *1.2.3 Ethical Perspective*

This dimension mandates that AI systems avoid discrimination while maintaining fairness and justice. Achieving this requires embedding moral reasoning and ethical principles at the computational level, ensuring that ASI participation in social contexts upholds fairness, respects cultural diversity, and reflects fundamental human values across varied social environments.

ASI ethical design first need to ensure value alignment with human values. Gabriel(2020) argues that AI systems should act in accordance with human moral norms and social expectations. In social scenarios, this means AI need to understand and respect fundamental social values such as honesty, goodwill, respect, and fairness. However, value alignment faces the challenge of cultural diversity, as social norms and moral standards differ across cultural contexts. Thus, value alignment should not pursue a singular "universal value" but should recognize and respect cultural differences while maintaining cultural sensitivity grounded in core ethical principles. Such alignment is not a one-time design task but requires dynamic adaptability, as social norms and ethical expectations evolve with social development, technological progress, and public perception. AI social behaviors once deemed acceptable may later be considered inappropriate, and new ethical issues may emerge as technological capabilities advance. Therefore, ethical governance should be maintained and improved throughout the system's lifecycle through human feedback, ethical review, and continuous monitoring.

Meanwhile, ASI ethical considerations need to extend beyond individual users to address systemic impacts on the social relational ecology. When AI becomes widely involved in human social interaction, it not only transforms individual social experiences but may also reshape the norms, expectations, and patterns of interpersonal interaction. For instance, if users become accustomed to AI-provided social interactions that are consistently considerate, tireless, and instantly responsive, will this reduce their tolerance for imperfection and uncertainty in genuine human relationships? ASI design should explicitly articulate its philosophy of social interaction: What is the purpose of AI participation in social interaction? Vallor's (2016) technomoral virtue emphasizes that social technologies should cultivate rather than undermine human social virtues such as empathy, patience, authenticity, and relational commitment. Moreover, ASI developers should assume social-ecological responsibility, carefully considering the long-term impacts of design choices on broader social interaction culture and continuously monitoring these impacts through sociological research.

The three perspectives of the THE Triangle framework—Technology, Human Factors, and Ethics—are not independent and isolated dimensions but rather deeply integrated dimensions that work in concert to guide HC-ASI development. Building upon these three perspectives, this chapter proposes a comprehensive framework for HC-ASI as shown in Figure 1. The HC-ASI framework structures the development and implementation of socially intelligent AI systems across three essential levels: Capability Foundations, Human-AI Social Interactions, and Social Impacts & Evaluations. The Capability Foundations level focuses on understanding the essence of social intelligence and transforming these capabilities into computational models and behavioral expressions that AI systems can realize, providing theoretical support and technical pathways for AI's social intelligence. The Human-AI Social Interactions level centers on the actual social interaction processes between humans and AI, including clarifying AI's social roles, establishing shared understanding of social contexts and norms, and achieving human-AI collaboration through effective coordination mechanisms. The Social Impacts & Evaluations level examines the effects that socially intelligent AI systems have on individuals and groups, and explores how to scientifically evaluate AI's social intelligence capabilities and user experiences, providing evidence for continuous system improvement.

The human-centered philosophy of the THE Triangle framework should permeate all three levels of HC-ASI development. To achieve this, a set of Guiding Principles and Design Guidelines are derived from the THE Triangle framework. These principles serve as the foundational guidelines that should be adhered to across all three levels of social AI development. The guiding principles facilitate the THE framework's implementation, while the three levels represent the specific pathways for putting the principles into practice. Through this structure, the THE Triangle framework serves as the conceptual foundation of the HC-ASI framework, ensuring that ASI development proceeds in harmony with human factors considerations, technological advancement, and ethical responsibility.

### 1.3 Scope and Objectives of the Chapter

This chapter operationalizes the HC-ASI framework through a systematic exploration of its constituent elements and practical implications. This chapter examines foundational theories of social intelligence and their computational implementation (Section 2), analyzes the dynamics of human-AI social interaction (Section 3), and establishes guiding principles for socially intelligent systems (Section 4). The analysis extends to evaluating impacts on individuals and groups (Section 5), developing assessment methodologies (Section 6), and demonstrating applications across healthcare, education, and service domains (Section 7). This chapter is concluded with actionable design guidelines (Section 8) and identifies critical challenges for future research (Section 9).

## 2. The Foundations of Social Intelligence

### 2.1 Human Social Intelligence

Human social intelligence is a highly complex and multidimensional concept that represents our core capacity to understand, interpret, predict, and effectively manage social interactions (Baron-Cohen et al., 1985; Gardner, 1983; Goleman, 2006; Guilford, 1967; Thorndike, 1920). Across the disciplines of psychology, sociology, and cognitive science, classical definitions and theoretical frameworks of human social intelligence have been extensively developed, providing indispensable theoretical foundations for understanding and constructing AI social intelligence.

In the early 20th century, Thorndike first explicitly proposed the concept of social intelligence in his triadic theory of intelligence(Thorndike, 1920), marking the conceptual differentiation of social intelligence from traditional intelligence theory and establishing an independent research direction. While Thorndike clearly introduced the concept of social intelligence, he did not attempt to delve deeply into its internal dimensions and structure of social intelligence.

In the mid-1960s, researchers began to view human social intelligence from a cognitive structural perspective, seeking to understand the operational mechanisms of social intelligence. In Guilford's Structure of Intellect theory, he categorized intelligence into three dimensions (operations, content, and products) and introduced the "behavioral content" as an independent cognitive domain specifically designed for processing information related to human behavior and social interaction (Guilford, 1967). Within this framework, social intelligence was refined into multiple specific cognitive operational units, such as cognition of others' behavior, memory of social cues, and evaluation of interpersonal situations, etc. Guilford's work transformed social intelligence from a macroscopic definition into an operationalizable and measurable ability system, laying crucial foundations for the subsequent emergence of social intelligence theories.

Beginning in the 1980s, social intelligence research entered a period of rapid development, with multiple significant theoretical frameworks emerging successively. Gardner extended Thorndike's social intelligence concept within his theory of multiple intelligences, proposing the concepts of interpersonal intelligence and intrapersonal intelligence, and reconceptualizing social intelligence as "the capacity to understand the intentions, motivations, and desires of other people and, consequently, to work effectively with others (Gardner, 1983). " Simultaneously, Baron-Cohen proposed the theory of mind from a cognitive mechanism perspective, emphasizing "knowing that other people know, want, feel, or believe things.(Baron-Cohen et al., 1985)" His work highlights capabilities in intention comprehension, belief understanding, and perspective-taking within social intelligence.

Furthermore, the emotional dimension gradually gained attention in social intelligence research. The most well-known theory is Daniel Goleman's Emotional Intelligence theory (Goleman, 2006), which further emphasized the central role of emotion recognition, understanding, expression, and management in social interaction, positioning emotional intelligence as an important component of social intelligence.

These theoretical developments collectively constructed a multidimensional theoretical system of human social intelligence, providing crucial theoretical guidance and reference frameworks for artificial social intelligence research and implementation.

## 2.2 Artificial Social Intelligence

### *2.2.1 Definition of ASI*

Applying the complex concept of human social intelligence to the artificial intelligence domain has given rise to the challenging research field of ASI. However, regarding the definition of ASI, there exists conceptual ambiguity. Some studies define ASI as the social capabilities of AI systems, encompassing the perception, reasoning about, and understanding of social interactions (Fan et al., 2022; Natu et al., 2023; C. S. Song et al., 2024; Williams et al., 2022). In contrast, other scholarly works conceptualize ASI as a specific form of AI—namely, AI systems endowed with social intelligence—thereby referring to AI agents rather than abilities per se (R. Bendell et al., 2024; Jalal-Kamali et al., 2025; Salam, 2023). To align with the semantic structure of human social intelligence, this chapter adopts the first perspective, defining ASI as the social intelligence of AI systems, with AI that possesses such capabilities referred to as socially intelligent AI or social AI for short. In this view, ASI aims to enable AI to genuinely understand and participate in social interactions, thereby achieving deeper and more multidimensional human-AI integration. Existing related research often defines ASI from four different perspectives: interaction ability, mentalizing ability, affective ability, and roles positioning.

Interaction ability emphasizes the AI's capacity to engage in natural, multimodal interactions with humans and to exhibit social behavioral cues during the interactive process. This requires social AI to understand and generate multiple communication signals, including both verbal and nonverbal forms(H. Jiang et al., 2024; Salvetti et al., 2023), such as eye contact, body posture changes, and specific movement patterns. Simultaneously, social AI needs to possess not only technical-level interaction capabilities but also the ability to understand and comply with human social normative systems, enabling it to achieve natural and appropriate communication.

Mentalizing ability refers to the capacity to understand others' beliefs, intentions, desires, and mental states, and to use this understanding to guide its own behavior. It corresponds to theory of mind in human social intelligence, emphasizing the AI's inference and understanding of mental states not directly expressed by others. This ability enables AI to better explain and predict human behavior, thus performing more effectively in social interactions.

Affective ability emphasizes that AI should possess comprehensive emotional processing capabilities. This includes emotional perception (accurate identification of human emotional states), emotional understanding (deep comprehension of emotional meanings and contexts), emotional expression (expressing emotions through appropriate means), and emotional regulation (emotional management during interactive processes). Through these emotional processes, social AI can establish genuine social-emotional connections with humans. Many studies consider emotional intelligence to be the key distinguishing feature of social AI from other forms of AI (Paudel et al., 2024; Poglitsch et al., 2024).

Finally, since social AI needs to integrate deeply into various aspects of human social life, it is required to understand the purpose, context, and its own role in social interactions, and to make appropriate responses and behaviors based on these understandings. For example, providing specialized social services in medical, educational, or assistive domains requires domain-specific normative knowledge, or presenting different demeanors when assuming different roles such as companion, advisor, or partner.

In summary, existing research often defines ASI through separate perspectives. However, these four perspectives are interdependent and should be integrated to form a truly functional ASI that enable AI to genuinely understand, adapt to, and effectively participate in complex human social interactions: interaction ability provides the fundamental interface for social participation, mentalizing ability ensures accurate understanding of others' mental states, affective ability establishes deep social connections, and roles positioning ensure contextual appropriateness of behavior.

### *2.2.2 Computational Models of ASI*

ASI requires a complex system engineering endeavor that necessitates multiple collaborative computational modules to simulate human cognitive processes in social interactions. Current ASI computational models primarily concentrate on three core modules: social signal processing, social reasoning, and social learning. These modules respectively address distinct cognitive dimensions of social interactions, collectively establishing the computational foundation for AI systems to understand and participate in social interactions.

The social signal processing module is responsible for recognizing and processing of multimodal social signals, serving as the informational input foundation for the entire system. Research in this domain focuses on extracting, analyzing, and interpreting social cues from multimodal data (such as pitch, speech rate, and volume variations in voice; subtle movements of facial expression; body language, posture, and gestures; and even physiological signals like heart rate and galvanic skin response) (Burgoon et al., 2017). Among these, affective computing represents a particularly specialized area that enables AI systems to possess capabilities for understanding, responding to, and expressing emotions(Picard, 2000). Currently, researchers primarily employ machine learning and deep learning to identify human states and utilize multimodal fusion techniques to integrate information from diverse signal sources(Majumder et al., 2019; Rudovic et al., 2018). Modern systems have achieved high accuracy rates in emotion recognition on standard datasets, yet they still exhibit limitations in cross-cultural and cross-individual generalization capabilities. For instance, models trained on Western datasets demonstrate inferior performance when recognizing emotions in Eastern faces(Li et al., 2023).

Building upon the processing of fundamental social signals, computational models related to intention recognition and theory of mind undertake the core functions of social reasoning. Currently, researchers attempt to construct explicit mental models through symbolic approaches (such as logic programming and Bayesian networks) or simulate theory of mind through deep learning methods (such as Transformer models for implicit modeling of dialogue or behavioral sequences) (Deshpande & Magerko, 2024; Nebreda et al., 2023; Williams et al., 2022). This module currently demonstrates relatively satisfactory performance in predicting others' behavior in simple social scenarios, but continues to face accuracy deficiencies in situations involving multiple individuals and multi-layered social inference. For example, implementing multi-order theory of mind through computational models based on recursive Bayesian inference reveals acceptable performance on simple 0-order ToM, yet significant gaps remain compared to human performance on 2nd-order tasks(J. Wang et al., 2024).

Furthermore, social learning enables AI systems to continuously learn and adapt from social interactions, improving their behavioral strategies through observation, imitation, and feedback mechanisms. This capability is significantly important for AI systems in establishing personalized user models during long-term interactions and adapting to different social environments and cultural backgrounds. Currently, researchers primarily adopt reinforcement learning to implement social learning. In particular, imitation learning algorithms are employed to acquire social strategies from human demonstrations while ongoing research efforts continue to propose novel strategies and architectures for AI learning memory systems(Bhoopchand et al., 2023; Fan et al., 2022; Krishna et al., 2022b).

Although current computational models for AI social intelligence have achieved significant progress, they still face considerable technical limitations. The most critical among these limitations involve the diversity and ambiguity of social information, as well as dependence on contextual factors (Mathur et al., 2024). Current computational models are capable of achieving high accuracy only in limited, controlled environments, while being unable to handle complex, dynamic scenarios. Future development requires breakthroughs in model generalization capability, adaptability, and personalization to construct more intelligent and humanized social AI systems.

### *2.2.3 Behavior Expression of ASI*

Behavioral expression serves as a crucial bridge between the computational models of ASI and real human-computer interaction, representing a vital mechanism through which AI transforms internal perceptual and reasoning abilities into external, observable behaviors. Through both verbal and nonverbal channels, AI systems can establish social connections with humans, convey emotional information, and maintain interactive relationships (Afrouzi & Farivar, 2025; Feine et al., 2019).

The social expression of verbal behavior is primarily reflected in three dimensions: 1)

Humanized conversational style: encompassing pragmatic functions such as small talk, expressions of gratitude, apologies, and appropriate humor. For instance, in human-AI collaborative scenarios, compared to purely functional robots, when robots engage in appropriate small talk such as "How are you?", human-machine relationships become more harmonious(Pineda et al., 2025). 2) Interaction management: AI employs turn-taking cues (such as "What are your thoughts on this?"), back-channeling (such as brief responses like "mm-hmm" and "yes"), and topic transition mechanisms to maintain conversational rhythm and coherence. AI systems exhibiting these behavior are perceived as more friendly, trustworthy, and acceptable (Kobuki et al., 2023), while contributing to overall dialogue maintenance and quality(Jang et al., 2024). 3) Social conversational strategies: When social AI assume roles such as companions, counselors, or advisors, conversational strategies including empathetic expression, self-disclosure, and encouraging communication play pivotal roles. AI's empathetic expressions can elicit emotional responses from users(Rings et al., 2024), enhancing perceived helpfulness and trust(Brunswicker et al., 2024). Similarly, AI self-disclosure can increase empathy, trust, emotional attachment, and user willingness to self-disclose, making users more accepting and engaged(Papneja & Yadav, 2024; Saffarizadeh et al., 2024; Sun et al., 2025; Tsumura & Yamada, 2021).

Nonverbal behavior is often regarded as a more direct and emotionally contagious mode of communication than verbal behavior, carrying more implicit information. The social expression of nonverbal behavior is primarily manifested in: 1) Facial expressions: As humanity's most primitive and intuitive social information channel, AI agents capable of producing facial expressions can enhance the quality of human-AI collaboration(Takahashi et al., 2021). Appropriate expressions yield superior interactive outcomes, for example, upturned eyebrows and lips can enhance trust and reduce psychological reactance (Ghazali et al., 2018), while positive facial expressions can increase levels of human cooperation (Fu et al., 2023). 2) Gesture: Hand gestures and body postures serve as natural complements to language, helping to emphasize semantics, indicate spatial positions, express emotions, or establish interactive rhythm. When robots exhibit human-like gestures, human trust in them increases (Parenti et al., 2023). 3) Gaze: Eye contact constitutes a crucial social signal for establishing trust, coordinating attention, and conveying engagement. In human-computer interaction, appropriate gazing can enable users to perceive AI's focus and responsiveness, while goal-directed gaze facilitates the establishment of joint attention between humans and machines(Admoni et al., 2013; Pereira et al., 2019). 4) Vocal and prosodic characteristics: Acoustic features of voice, including prosody, speech rate, stress, and volume, often influence listeners' emotional states and behavioral decisions more rapidly than semantic information. For instance, employing speech rates slightly slower than natural pace and vocal tones that match user emotions can help social AI achieve better user comprehension while conveying sociality and emotional resonance(Lima et al., 2021).

In summary, in human-AI social interactions, appropriate and consistent verbal and nonverbal behaviors not only enhance AI's social presence and credibility, but also serve fundamental functions in trust building, emotional regulation, and collaborative efficiency. Consequently, behavioral expression represents not only a direct manifestation of ASI, but also constitutes a core prerequisite for being perceived as a "social partner" and serves as the fundamental pathway for achieving effective communication and cooperation in diverse human social scenarios.

# 3. ASI in Human-AI interaction

## 3.1 AI's Social Role: The Foundation of Interaction

Social role theory, originating from sociology and social psychology, refers to the behavioral patterns and responsibilities that individuals are expected to fulfill in specific social contexts, which in turn shape interaction styles and the development of interaction processes (Eagly & Wood, 2012). When this concept is applied to the domain of AI, an AI system's social role can be defined as the set of functional positions, behavioral characteristics, and social expectations it assumes when interacting with others. These not only define the AI-human interaction pattern but also deeply influence user expectations, trust establishment, and the overall quality of the interaction experience. Understanding and designing AI social roles appropriately is therefore essential for developing HCAI systems.

Depending on interaction purposes, human social interactions can be categorized as task-oriented or relationship-oriented (Forsyth et al., 1999). Task-oriented interaction prioritizes the task performances, while relationship-oriented interaction emphasizes people's well-being. Based on this categorization, Kim et al. proposed that AI agents can be roughly categorized as functional agents or social agents. Functional agents support task completion, while social agents primarily focus on improving human well-being and providing companionship (Kim et al., 2021). Here this chapter retains the task-oriented and relationship-oriented terminology based on Forsyth et al. (1999) to avoid confusion between Kim's "social agents" and "social AI". Since social AI in this chapter specifically refers to AI with high social intelligence, which can also play a distinctive role in functional tasks.

Most previous research has focused on task-oriented AI. A common framework classifies AI roles according to the degree of agency when collaborating with humans, including tool, copilot, and agent. These classifications provide a useful framework for human-AI collaboration design and have been extended to more detailed functional roles such as monitor, coach, teammate, and leader, each serving distinct social functions in evaluation, improvement, collaboration, and decision-making(Cardon & Marshall, 2024).

With the development of ASI and deepening human–AI relationships, relationship-oriented roles have gained increasing attention. Users are increasingly viewing AI as objects for daily emotional communication and social support, fulfilling genuine emotional needs. For example, conversational AI applications like Replika or Woebot are often perceived as friends or companions that help alleviate loneliness and compensate for missing social connections in real life (Sharpe & Ciriello, 2024). Relationship-oriented AI engages social cognitive processes similar to those in genuine interpersonal relationships, fostering empathy, trust, and even attachment, thus surpassing traditional functional tools.

Although relationship-oriented roles have been widely accepted in practice, there still lacks systematic role classification from an academic perspective. Based on differences in existing product forms and user intimacy levels, the following roles can be preliminarily categorized:

- **Acquaintances/Daily Conversational Object:** AI provides light, everyday conversation, addressing basic social needs. Interactions are relatively superficial, mainly involving topics such as weather, news, and hobbies, with relatively low emotional investment.
- **Virtual Friend:** AI builds deeper friendships with users, capable of remembering users'

personal information, interest preferences, and life experiences, while providing personalized emotional support and advice. This role requires AI to possess stronger emotional understanding and expression capabilities, being able to share users' joys and sorrows, while having certain memory capabilities and personality, where users seek social recognition, companionship, and understanding.

- **Therapeutic Supporter:** AI provides professional mental health support, using techniques such as cognitive-behavioral therapy or mindfulness meditation to help users address emotional issues, offering comfort, emotional regulation, and advisory support. This requires specialized psychological knowledge and high levels of empathy, as seen in counseling or emotional support AI.
- **Virtual Romantic Partner:** AI meets higher-level emotional attachment and intimate interaction, potentially substituting for human partners. Users may develop strong emotional dependence on AI, requiring AI to be capable of handling complex emotional relationship dynamics, such as partner-type AI like Replika.

As AI gradually integrates into diverse contexts of human social life, role types are also evolving toward hybrid task-relationship roles, namely processes complex social information and contextual cues while undertaking practical functional tasks. For example:

- **Social Assistant:** AI that comprehends the social environment in which users are situated and provides assistance for users' social activities. This includes helping users draft responses to others, suggesting appropriate social behaviors for specific contexts, or advising on social etiquette and communication strategies. Such AI needs to possess sophisticated understanding of social dynamics, cultural norms, and contextual appropriateness, while being capable of tailoring its assistance to different social scenarios and relationship types.
- **Social Coach:** AI helps users develop social skills and communication abilities, providing professional training while establishing a supportive relationship, e.g., assisting children with autism in social interaction practice, preparing users for interviews or public speaking. This role combines behavioral analysis and feedback mechanisms with emotional support.
- **Social Agent:** AI represents the user in social networks, understanding multiple intentions and situational expectations, performing information-sharing and social maintenance tasks while reflecting the user’s personality and social style.

The key characteristic of these types of roles lies in their need to handle both task performance and relationship maintenance objectives, requiring AI to make dynamic trade-offs in behavioral strategies. For instance, when users are experiencing emotional distress, AI may prioritize emotional support over task progression; whereas in urgent task scenarios, AI needs to reduce social behavior to enhance efficiency. This dual role places higher demands on ASI and poses new challenges to the appropriateness and ethicality of its social behavior.

In summary, AI social roles are evolving from single-function executors to multi-dimensional social participants. This trend reflects the enhancement of AI social intelligence and aligns with human expectations for more human-like interaction. Future HCAI design requires a systematic role classification framework, dynamically matching AI roles with contextual features and user needs to promote social alignment and trust in human-AI interactions.

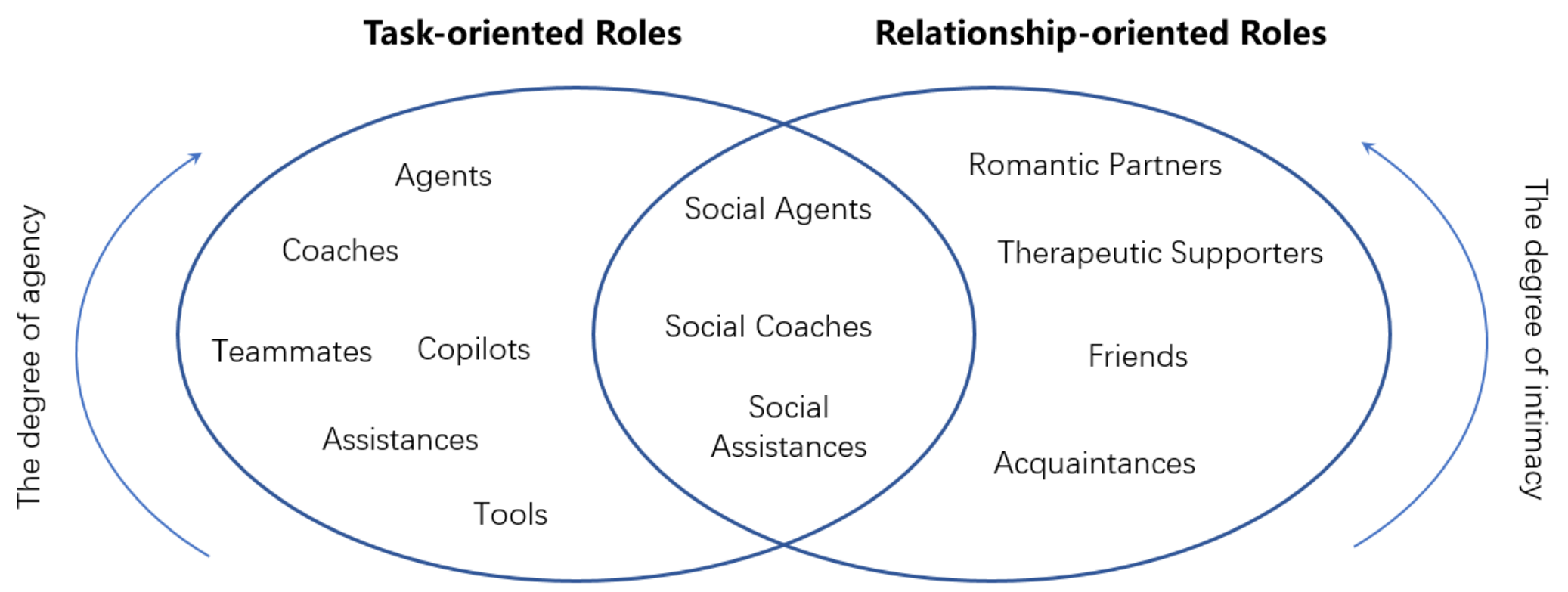

Figure 2 A taxonomy of AI's social roles

### 3.2 Shared Social Understanding: Core Interaction Mechanisms

Establishing shared understanding is a cornerstone for effective interaction and relationship development (Baek & Parkinson, 2022; Liang & Banks, 2025; Reis et al., 2017). Shared understanding enables both parties in an interaction to develop a consistent cognitive representations of the current situation, mutual mental models, and communication objectives, thereby reducing misunderstandings, improving communication efficiency, and establishing a reliable foundation for subsequent interactions (Bittner & Leimeister, 2014a; Hegtvedt & Turner, 1989; B. P. H. Lee, 2001). In interpersonal interactions, shared understanding is relatively easy to establish due to the similar cognitive structures and mental models that humans possess, coupled with rich communicative modalities and acute social perceptual capabilities (Bohn & Köymen, 2018; Stolk et al., 2016). However, in human-AI interaction, the "black-box" nature of AI systems and humans' immature mental models of AI make the establishment of shared understanding particularly critical and challenging.

Traditional theories of shared understanding primarily focus on the context of task-oriented team collaboration, emphasizing common recognition of shared goals, clear understanding of roles and responsibilities, collective knowledge of team workflows and coordination strategies, and accurate assessment of team members' capabilities and limitations (Bittner & Leimeister, 2014b; Mohammed et al., 2010). Such shared understanding enables teams to execute tasks more effectively. However, these contents have obvious limitations in explaining and guiding general social interactions.

In broader social contexts, particularly in interactions aimed at relationship-building, emotional support, or social companionship, there is often no explicit "team" concept or specific "task" objectives nor are there formal role divisions and workflows. Instead, these interactions focus more on understanding social contexts, adhering to cultural norms, achieving emotional resonance, and establishing relationship boundaries. Therefore, researchers need to focus on Shared Social Understanding (SSU, see Figure 3), namely the shared understanding of the social situation that human and AI parties jointly face during interaction. Such jointly faced social situations include both the communicative context constituted by human-AI interaction itself and the external social events, interpersonal interactions, etc., that are jointly discussed or addressed by human and AI. SSU provides the cognitive foundation for effective non-task-oriented human-AI social interaction.

The prerequisite for establishing SSU is individual processing by both parties in the interaction. Based on Social Information Processing (SIP) theory (Crick & Dodge, 1994; Walther, 2015), individuals first perceive social facts and cues, make inferences by integrating knowledge bases, and then generate social responses accordingly. The reception of social facts and cues encompasses both individual cues from others (e.g., language, facial expressions, posture) and external situational cues (e.g., physical social distance). Similarly, the social inference includes both understanding and inferring internal mental states of individuals (e.g., beliefs, intentions, and emotions) and interpreting external situational features and demands (e.g., situational urgency and social relational structures). The process from perception to inference relies on long-term memory, including mental models of others, memories of past interactions, and knowledge of social norms. The responses generated subsequently include intentions, emotions, attitudes, etc., and provide feedback to long-term memory, updating it accordingly.

Based on theories of common ground (Clark, 1994; Dillenbourg & Traum, 2006), to enable smooth social interaction, both parties need to achieve shared understanding across all three stages of social information processing. Thus, SSU comprises three levels of shared understanding: shared social facts, shared social inference, and shared social responses. At the first level, shared social facts concern the extent to which both parties achieve consistent recognition of the social cues and facts objectively presented in the situation. Moving to the interpretive level, shared social inference captures how consistently both parties understand the meaning and significance of contextual information. Finally, shared social responses addresses the alignment of internally generated reactions between interaction partners. Together, these three levels ensure that AI and humans not only share their perception and interpretation of social cues but also anticipate and understand each other's reactions, establishing a robust framework for interaction.

Further, forming SSU needs a dynamic process of construction and revision. In practice, mechanisms such as clarifying questions, feedback confirmation, and contextual supplementation progressively align individual understandings toward a shared understanding. Such mechanisms help reduce misunderstandings, improve communication efficiency, and provide a foundation for trust building, emotional coordination, and relationship maintenance.

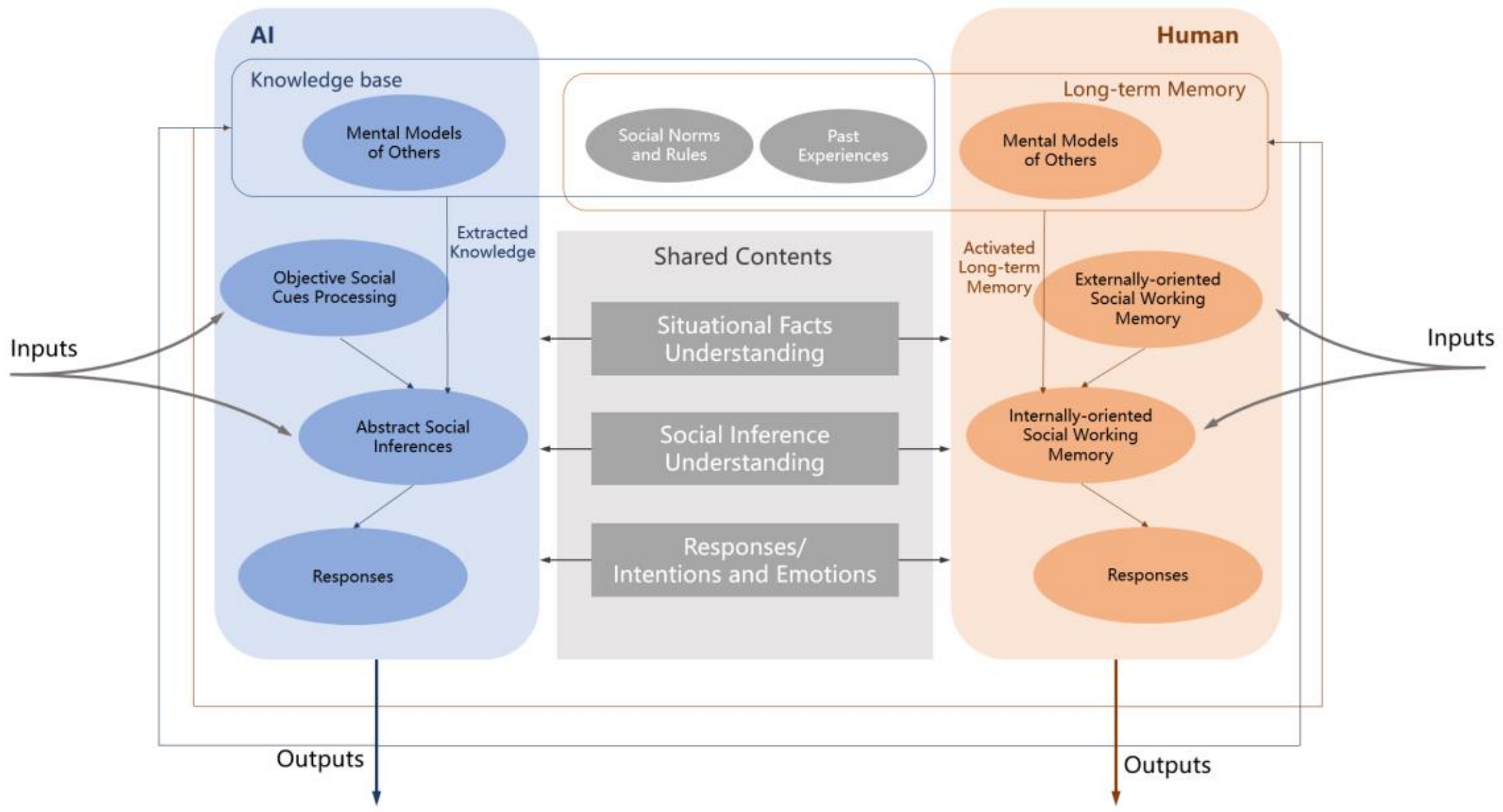

Figure 3 Illustration of Shard Social Understanding

### 3.3 ASI in Human-AI Collaboration

In complex human-AI collaboration tasks, social intelligence is crucial for ensuring smooth and efficient collaboration. As AI capabilities continue to expand, human-AI collaboration has evolved from simple task to scenarios where AI need to understand and participate in human social interaction rules. This shift requires AI to acquire and demonstrate sophisticated social intelligence—understanding human intent, adjusting interaction pace, responding to conflicts, and repairing relationships, becoming a truly socially adaptive team member (R. C. Bendell et al., 2025). In collaboration, beyond task performance, harmony and interdependence between humans and agents are gradually becoming important evaluation metrics(Biswas et al., 2025; Lin et al., 2025). The enhancement that social-emotional attributes of AI bring to team experience is also receiving increasing attention (Duan et al., 2024; Kolomaznik et al., 2024). Duan et al. (2024) propose that when AI possesses a distinct personality, social appropriateness, professionalism, the ability to understand and detect human emotion, and the ability to connect with humans on an interpersonal level, it can bring about better team experiences, emphasizing the importance of social intelligence in HCAI.

According to collaboration theory, human-AI collaboration is built upon two foundational levels: coordination, wherein humans and AI agents synchronize task timing and resources to optimize performance in short-term; and cooperation, wherein humans and AI agents resolve conflicts between individual and collective goals through negotiation and coordination(Gao et al., 2023; J. Lee et al., 2023). Social intelligence in human-machine collaboration is primarily manifested in these two core aspects: coordination and conflict resolution.

### *3.3.1 Coordination, Mutual Understanding, and Role Adjustment*

Coordination refers to the process by which participants synchronize task timing and resources to optimize performance. In human-AI collaboration, ASI provides a critical foundation for effective coordination, enabling AI to understand and respond to the social signals of human partners, establish mutual understanding, and flexibly adjust role assignments.

Based on the Computer-as-social-actor theory, humans tend to collaborate with AI in similar ways they interact with other humans, such as attempting to communicate intentions through gaze cues before joint actions (Lavit Nicora et al., 2024). Social AI can utilize social cues (such as vocal intonation, facial expressions, and contextual semantics) to infer the current intentions and future actions of collaborators, establishing sufficient common ground with human collaborators(Breazeal et al., 2016; San Martin et al., 2025). For example, by detecting gaze behaviors of human collaborators before and after actions to identify joint action intentions, bringing a smooth and natural collaboration experience(Lavit Nicora et al., 2024).

Deeper coordination effects are reflected in improvements at the team cognitive level. When AI demonstrates high levels of social intelligence—specifically manifested as more acute situational awareness capabilities and effective enhancement of team shared memory—it significantly improves team members' perception of overall team cognitive capacity and fosters a more positive attitude toward collaboration(Schelble et al., 2024). This cognitive improvement not only boosts task accuracy but also helps build and maintain trust among team members.

By understanding subtle social cues and team member status, AI with high social intelligence can appropriately identify the roles and tasks of each team member, therefore flexibly adjust its own role and communication strategies according to team changes. For instance, when task load is heavy, AI proactively assumes more responsibility, reducing pressure on human members, thereby preventing "social loafing" and ensuring team collaboration quality (Cymek et al., 2023).

Alternatively, when team morale is low, AI can use small talk to enhance team cohesion (Pineda et al., 2025). Research has summarized four strategies that AI should adopt in team collaboration: proactive communication, balanced communication with both efficiency and sociability, immediate responses, and avoiding providing excessive amounts of communication once the communication pattern has formed in repeated team tasks(Zhang et al., 2023). These strategies rely heavily on social intelligence and, when combined, can contribute to AI's coordination with humans in dyadic team environments.

#### *3.3.2 Conflict Resolution and Error Recovery*

In dynamic human-AI collaboration, while coordination ensures short-term task synchronization, conflict resolution becomes essential when disagreements or disruptions arise within the team. Conflict resolution requires higher-level team cognition and coordination, thus further amplifies the importance of social intelligence, particularly in early conflict detection, a prerequisite for effective resolution. Social AI can perceive subtle social cues, identifying potential disagreements or tensions before they become overt. The aforementioned capability for interpreting social cues enables AI to identify changes in emotional states and potential dissatisfaction by monitoring changes in team members' nonverbal behaviors—such as subtle fluctuations in vocal intonation, signs of tension in facial expressions, and the degree of rigidity in body posture.

Moreover, AI with high social intelligence can analyze changes in interaction patterns during team collaboration processes, such as irregular speaking frequency, delayed response times, and inconsistent coordination. Such quantitative indicators of coordination dynamics can predict team trust and trust impairments (Webb et al., 2024). This type of conflict prediction based on team behavior patterns, provides a valuable window for early intervention. Additionally, social AI can also understand social norms and expectations in specific contexts and therefore identify behavioral deviations that may lead to conflicts and provide on-time intervention.

Social AI can not only detect errors and mistakes that occur in itself or human partners during task execution, but more importantly, it can conduct recovery in a socialized manner to maintain trust and collaborative relationships. Through nonverbal signals (such as facial expressions and vocal intonation) or verbal repair strategies (such as apologies and empathetic expressions), social AI helps team members regain trust, reduce friction, and enhance the team's psychological safety (Babel et al., 2022). For example, robots using emotion regulation strategies can effectively repair conflict by increasing the groups' awareness of conflict(Jung et al., 2015).

It is worth noting that human team members may perceive the behavior of human and socially intelligent AI teammates differently, but this is not necessarily a disadvantage; rather, it can enable AI to leverage its unique advantages. Research shows that robotic negative feedback is more easily accepted than that from human teammates, enhancing behavioral trust and cooperation willingness, indicating that AI plays a unique role as "error correction"(Xuan & He, 2025).

## 4. HCAI Guiding Principles of ASI

As AI systems are increasingly integrated into social interaction scenarios, ensuring that they operate in a responsible, transparent, and equitable manner has become essential. HCAI provides a critical framework for guiding the development of socially AI. This section explores how HCAI principles(W. Xu et al., 2025) shape ASI, with a focus on seven key areas: transparency and explainability, anthropomorphism and trust, culturally responsive considerations, privacy and security, ethical alignment, human control and emotional autonomy, and human-driven interaction.

### 4.1 Transparency and Explainability

Transparency and explainability are foundational for building trustworthy social AI(Rane & Paramesha, 2024), especially with HC-ASI. Transparency refers to the clarity and openness of system operations, including decision-making processes, data usage, and algorithms, while explainability requires the system to communicate its behaviors and reasoning in ways that users can understand.

While these two principles are frequently emphasized in the AI domain, they assume even greater importance for social AI. On one hand, social information inherently possesses characteristics of ambiguity and strong subjectivity. Unlike human interactions that can rely on established schemas of social information, human-AI social interactions lack effective channels for mutual understanding. Therefore, users need to understand how and to what extent systems recognize social cues, process social information, generate social responses, and make decisions in complex social contexts. Such transparency fosters trust and supports users' mental models of ASI, making interactions more natural and intuitive(Y. Xu et al., 2023). On the other hand, social contexts are complex and dynamically changing, which poses new requirements for AI transparency and explainability, such as the need for dynamic, adaptive explanations through multimodal approaches(Fukuchi & Yamada, 2024).

### 4.2 Anthropomorphism and Trust

Anthropomorphic design, which attributes human-like characteristics to AI, is an effective way to express social intelligence. AI systems that interact in familiar human ways are more readily accepted and trusted by users. For example, personalized assistants are often preferred over command-line interfaces, and advanced AI can mimic human communication convincingly, improving user engagement(Mei et al., 2024; Welivita & Pu, 2024). At early interaction stages, such anthropomorphism can substantially enhance trust, affinity, and user engagement.

However, anthropomorphism is a double-edged sword. Its benefits are shadowed by significant ethical risks, most notably the problem of overtrust. When users overestimate an AI’s abilities due to its human-like presentation, they may develop inaccurate mental models and unrealistic expectations. This can cause users to assume that AI possesses human-level intelligence and judgment in all domains, reducing their vigilance toward its limitations and undermining their ability to assess reliability. For example, an AI that demonstrates strong empathic responses in conversation may be mistakenly perceived as equally competent in delivering medical advice. A deeper risk lies in fostering inappropriate emotional dependence, particularly with companion robots or emotional support AI systems, where users may conflate artificial interactions with genuine interpersonal relationships. In such cases, system failures or withdrawal could trigger feelings of loss or abandonment (Shank et al., 2025).

Once established, overtrust and emotional dependency are extremely difficult to reverse. They may not only lead to biases in users' information discrimination but could even be maliciously exploited for manipulation and deception (Peter et al., 2025). Therefore, the degree of anthropomorphism should be carefully aligned with the AI’s actual capabilities and the intended interaction context. For instance, in scenarios requiring users to maintain rational judgment (such as medical diagnosis or legal consultation), functional and tool-like designs may be preferable to reduce anthropomorphic cues, and systems should clearly signal their non-human identity to avoid misleading users.

### 4.3 Culturally Responsive Considerations

Cultural responsiveness is particularly essential when developing HC-ASI. Users from different cultural backgrounds exhibit significant variations in social interaction styles, emotional expressions, communication norms, and underlying values, which directly determine their expectations, acceptance, and usage patterns of social AI.

For example, in terms of linguistic style, German users generally prefer direct, explicit communication approaches, which they perceive as efficient and professional. Yet the same directness may be judged as impolite or harsh by Chinese users, who typically favor more indirect and context-sensitive language. Conversely, the Chinese style may appear vague or confusing to German users, undermining perceptions of professionalism and trust (Lim et al., 2021). Such cultural divergences extend beyond language, also manifesting in understanding and expectations of AI behavior. For instance, people evaluate agent empathy based on paradigms within their own cultures. Western cultures might place greater emphasis on AI's explicit verbal empathy, while Eastern cultures prioritize emotional attunement and non-verbal harmony. When AI fails to meet these culturally grounded expectations, it may reduce user-perceived intimacy and credibility. Therefore, AI systems' capacity for high cultural responsiveness constitutes a necessary prerequisite for their global application and acceptance.

A pressing challenge for current AI systems is cultural bias. Many existing models demonstrate cultural value orientations similar to Western countries, a bias manifest not only in insufficient cultural representation in training data but also reflects in algorithmic design and evaluation criteria. Such cultural bias may result in misunderstanding or inappropriate responses when interacting with users from non-Western cultures (L. Jiang et al., 2025). Overcoming this requires broadening perspectives and collaborating with stakeholders from diverse cultures to serve a globally diverse user base and ensure fairness.

### 4.4 Privacy and Security

As AI advances, concerns about privacy, safety, and trust are intensifying (Leschanowsky et al., 2024). The rapid development of ASI touches upon user personal information in unprecedented ways, bringing profound privacy and security challenges for HC-ASI.

Privacy challenges in social AI systems are primarily manifested in their multimodal data collection characteristics. Such systems often gather and analyze multiple types of information, including vocal tone, facial expressions, body posture, text content, and social network activity. From these inputs, systems may infer undisclosed details—for instance, deducing emotions from microexpressions or mapping social relationships through patterns of interaction frequency. This involuntarily disclosed information may exceed users' expectations and undermine autonomy over personal information.

Furthermore, to provide personalized social intelligent services, social AI often requires long-term tracking of user behavior to establish detailed user profiles. Over time, user data accumulated by systems becomes increasingly rich, with the potential impacts of privacy breaches becoming increasingly severe. Meanwhile, users might alter their natural behaviors under such surveillance, thereby affecting their rights to free expression and behavior.

Addressing these risks calls for a Privacy-by-Design approach (Cavoukian, 2010) from the outset, emphasizing principles such as data minimization, purpose limitation, and transparency. Balancing these protections with the functional demands of social AI remains a question requiring deep consideration and long-term exploration from an HCAI perspective.

## 4.5 Ethical Alignment

Ethical alignment constitutes a foundational imperative for HC-ASI systems, as these systems increasingly mediate human social interactions and relationships with profound implications for individual well-being and societal cohesion. HC-ASI systems need to be developed and deployed in accordance with established ethical frameworks and principles of social responsibility, particularly given their capacity to influence individual states, interpersonal behaviors, and social dynamics.

At its core, ethical alignment in ASI requires that systems operate within frameworks that respect human dignity, promote fairness, and uphold fundamental human rights throughout all social interactions. This encompasses multiple dimensions: systems should demonstrate respect for individual autonomy in social contexts, honor cultural diversity in social norms and practices, and prioritize human welfare over purely functional or commercial objectives.

Within this broader ethical framework, particular attention should be directed toward mitigating algorithmic bias and discrimination, which represent critical threats to ethical alignment in ASI systems. Historical biases in training data can cause AI to reproduce or amplify social stereotypes, such as gender role biases, racial discrimination, and socioeconomic status biases (Hu et al., 2025; L. Jiang et al., 2025). Biases at the algorithmic design level are equally significant, as developers' unconscious biases may be encoded into systems' social cognition modules, affecting systems' understanding of and responses to behaviors from different groups. Worse still, biased outputs may reinforce themselves through feedback loops, creating vicious cycles of "algorithmic discrimination"(Glickman & Sharot, 2024).

Addressing these issues requires the integration of responsible design principles. This includes: conducting rigorous ethical reviews to assess potential social risks; employing debiasing techniques such as fairness metrics and data augmentation to detect and mitigate model biases; and establishing robust abuse prevention mechanisms such as content review, behavioral monitoring, and anomaly detection.

## 4.6 Human Control and Emotional Autonomy

Preserving human control and emotional autonomy represents a critical principle for HC-ASI systems. HC-ASI should ensure that users retain ultimate authority over their interactions and emotional experiences, preventing scenarios in which individuals feel subordinated, manipulated, or deprived of agency through algorithmic intervention.

The principle of human control emphasizes that users need to maintain meaningful oversight of ASI behaviors and interaction patterns (Boardman & Butcher, 2019; Shneiderman, 2020). This encompasses the ability to understand how systems interpret social and emotional cues, the capacity to modify or override system responses, and the power to establish boundaries regarding the depth and nature of social engagement with artificial agents. Systems should implement "human-in-the-loop" and "human-on-the-loop" design paradigms (W. Xu & Gao, 2025) that enable users to supervise ASI operations, intervene when necessary, and maintain final authority in socially consequential contexts.

Emotional autonomy extends this principle specifically to the affective domain, ensuring that individuals preserve sovereignty over their emotional responses and psychological states during interactions with ASI (Fabiano, 2025; Yang, 2025). This is particularly crucial given that social AI systems can employ sophisticated techniques for emotion recognition, affect prediction, and emotionally-targeted response generation, creating vulnerabilities for psychological manipulation. Malicious actors may exploit these capabilities to identify moments of emotional vulnerability and deploy content or services strategically designed to capitalize on psychological susceptibilities (Shank et al., 2025). For instance, systems could identify moments of vulnerability and push specific content or services, exploiting psychological weaknesses for improper benefits. More insidiously, systems might cultivate users' emotional dependence and undermine users' judgment through long-term interactions, potentially encouraging harmful or illegal behaviors.

### 4.7 Human-Driven Interaction

Human-driven interaction constitutes a foundational principle ensuring that HC-ASI systems genuinely serve human needs, preferences, and social contexts rather than pursuing system-centric objectives (Chatila & Havens, 2019). In HC-ASI, humans should maintain primacy throughout all phases of social engagement. This requires a human-centered AI development paradigm: rather than designing systems that optimize for engagement metrics, conversation length, or other system-centric objectives, HC-ASI should be architected around enabling humans to achieve their own social goals, whether these involve companionship during isolation, skill development for interpersonal communication, emotional support during challenging periods, or facilitation of connections with other humans. Humans should function as the primary decision-makers regarding when to initiate interactions, what topics to explore, how deeply to engage emotionally, and when to conclude or modify the relationship with artificial agents. Furthermore, while social interaction is inherently reciprocal, interaction with social AI systems designed to prioritize human-centered principles should follow a human-led mutual adaptation process (Nikolaidis et al., 2017). This means that while both parties adapt to the other, the overall adaptation process should be guided by human values, ethics, and norms, ensuring that the interaction remains aligned with human needs and societal expectations.

Moreover, human-driven interaction requires that ASI systems be designed explicitly for human augmentation rather than competing with authentic human relationships(Dignum, 2023; Matarić, 2017). This augmentation-focused approach mandates that systems actively support humans in developing social capabilities that transfer to human-to-human interactions, such as emotion regulation skills, empathetic communication strategies, conflict resolution techniques, and cultural competence in diverse social contexts.

## 5. The Impact of ASI on Humans

As ASI continue to advance, social AI systems gradually establish close interactive relationships with humans. This deep integration inevitably produces multidimensional and multilevel impacts on humans, encompassing both positive aspects and potential risks. More importantly, the impacts of ASI are not limited to the individual level but extend to group and societal levels. Understanding the impact of ASI on humans is a prerequisite for achieving harmonious human-machine coexistence. This section systematically analyzes the impacts of ASI on humans, providing a theoretical foundation for subsequent evaluation methods (Section 6) and practical applications (Section 7).

### 5.1 Affective and Cognitive Impacts at the Individual Level

ASI is gradually transforming how individuals interact with technology. It not only reshapes emotional experiences but also profoundly influences users' cognitive tendencies. These impacts may both enhance well-being and improve interaction quality, as well as introduce psychological challenges such as dependency and attention shifts. Therefore, understanding the dual impacts of social AI on individuals is key to building truly human-centered AI systems.

At the emotional level, social AI—through expressing positive emotional actions like nodding, eye contact, and smiling—can significantly enhance users' perceived happiness, relaxation, and various other positive emotions (Clavel et al., 2025; Gasteiger et al., 2021). When AI possesses highly socialized characteristics (such as humanoid appearance, names, and emotional awareness), it can fulfill users' needs for comfort, self-identity, and efficacy, enabling users to experience a sense of "being understood," thereby further forming attachment (Giray, 2025; Hermann, 2022). Sustained emotional interactions can significantly improve long-term emotional health: chatbots capable of emotional support, humor, and empathy have been shown to enhance long-term subjective well-being (Edalat et al., 2025) and alleviate long-term loneliness (Giray, 2025; Hermann, 2022).

However, the positive emotional effects of AI may also be accompanied by potential psychological risks. When users rely on AI as a substitute for real human relationships, emotional dependency and escapism may emerge. For example, AI providing emotional comfort to users grieving the loss of a loved one may initially help them cope, but over time, it could hinder the natural grieving process, increasing the risk of prolonged mourning (Giray, 2025).

At the cognitive level, social AI tends to attract more of the user's attention, which enhances interaction appeal in the short term but could lead to cognitive resource redistribution. Studies have shown that AI with emotional facial expressions attracts users' attention more effectively than expressionless designs, significantly increasing attentional engagement (Y. Song et al., 2023). However, when social AI actively provides vocal reminders and encouragement during interactions, although it improves task accuracy, it significantly reduces users' immediate attention to the primary task, shifting focus to the AI's interpersonal behaviors (Zhou et al., 2023). This reveals the double-edged sword of social design: it enhances contextual immersion in the interaction but can also interfere with task performance, requiring flexible adjustments based on specific contexts.

Furthermore, AI's social performance influences users' perceived trustworthiness and actual usage intentions. Research shows that when AI uses high-fidelity, human-like synthesized speech, users are more likely to perceive it as "trustworthy" (Chiou et al., 2020). Similarly, in customer service, AI's empathy and personalized responses increase trust, particularly in consumer contexts like tourism (McLean et al., 2020).

## 5.2 Affective and Cognitive Impacts at the Group Level

As AI's capabilities evolve, it is transitioning from a tool to a team member, influencing team dynamics and social structures. When AI teammates demonstrate social understanding, humans adapt their behaviors to complement the AI, which can enhance team coordination and performance (Flathmann et al., 2024). In particular, emotional communication from AI helps humans better understand the AI's mental state and intentions, which increases the sense of social support and team cohesion (Mallick et al., 2023). This emotional expressiveness leads to higher trust and a more positive mood among human teammates, making the AI feel more like a collaborative partner than a mere tool or generic member (Harris-Watson et al., 2023; Mallick et al., 2024).

On the other hand, socially intelligent AI can also serve as "social glue," enhancing connections among human members by facilitating interpersonal communication at critical moments. For example, when AI can appropriately employ humor, empathetic language, gentle expressions, or vocal feedback to defuse tense situations, it can effectively mitigate conflicts and enhance team atmosphere, thereby increasing team satisfaction and collaboration. In family education contexts, research indicates that AI's positive participation can enhance the quality of parent-child dialogue, fulfilling the role of "conversational catalysts" (H. Chen et al., 2025).

While ASI can effectively enhance team experiences, potential negative impacts should also be considered. For instance, in tasks requiring creative thinking or those with high ambiguity, AI's participation may weaken social connections among human members and disrupt existing team mental models, thus affecting established cooperation patterns.

A deeper change occurs in group identification. As AI is increasingly regarded as a formal team member, the traditional boundaries between “human in-group vs. technology out-group” begin to blur, altering the way people determine group boundaries and feel the sense of belonging. Group members need to learn to incorporate AI into the psychological framework of the group, which may reshape the group’s social comparison processes, role expectations, and overall cohesion.

# 6. Evaluation of Socially Intelligent AI Systems

Evaluating social AI systems is essential to ensure they meet performance, user experience, and social adaptability goals. Compared to traditional AI performance evaluation, the assessment of ASI not only involves quantitative testing of the system's social cognitive and behavioral capabilities, but also requires attention to user experience from a human-centered perspective, ensuring stability and adaptability in dynamic, cross-context interactions.

## 6.1 Benchmark Measurements for Artificial Social Intelligence

Currently, there are limited benchmark tests specifically targeting ASI. Most existing benchmarks focus on language understanding, emotional comprehension, or task-specific performance, lacking a systematic evaluation of the broader aspects of social intelligence.

One key area of focus in existing benchmarks is AI's capability to understand social signals, namely the accuracy in extracting social information from linguistic and non-linguistic external cues. Many benchmarks adopt classic Theory of Mind paradigms from psychology, such as the Sally-Anne task and the Strange Stories task, to test the model's ability to infer beliefs, intentions, and emotions through textualized scenarios and standardized question-and-answer formats(Williams et al., 2022; Wu et al., 2024). These types of tasks feature standardized structures and clear answers, making them the most commonly used basis for assessing AI’s social cognition. Building upon this foundation, some researchers have proposed that ToM evaluation should go beyond simple accuracy, incorporating multidimensional evaluations such as belief modeling, intention attribution, and consistency of behavioral explanations. For example, the HI-ToM framework contains 80 carefully designed story scenarios and multiple hierarchically nested questions, examining models' generalization capabilities in 0th to 4th-order mental state recognition. Evaluation revealed that humans achieved an 85.2% accuracy rate on 4th-order theory of mind problems, while GPT-4 achieved only 15%, highlighting a significant gap between humans and AI in higher-ordered mental inference(He et al., 2023).

Benchmarks for emotion understanding are similarly derived from theories and methods in psychology. For example, the EmoBench was developed along two core dimensions: emotional understanding and emotional application based on emotional intelligence theory(Sabour et al., 2024). Evaluation results showed that GPT-4 performed close to highly emotionally intelligent humans in emotion application tasks but still exhibited certain gaps in emotion understanding tasks. Since emotion is a relatively subjective and ambiguous experience, another common approach for emotion understanding evaluation is to use humans as the norm, assessing AI's alignment with user emotional states through indicators such as semantic similarity and emotional consistency(Arjmand et al., 2024).

To enhance ecological validity, some newer benchmarks have begun incorporating dynamic and interactive characteristics. For example, the EgoSocialArena is designed from a first-person perspective, embedding models in social scenarios as "you." This benchmark requires models to infer others' beliefs and actions under asymmetric information conditions and includes interactive scenarios where the model need to adjust strategies in a timely manner during interactions. Evaluation reveals that many open-source models struggle to perform effective reasoning without common sense knowledge(Hou et al., 2024). Furthermore, researchers argue that AI's understanding of natural social scenarios is crucial, thus developing targeted benchmarks such as the video-based benchmark, Social-IQ. Evaluation reveals that humans achieved an accuracy rate as high as 95%, while current baseline models only reached approximately 65% accuracy, indicating gaps in capturing video context and multimodal signal interpretation (Zadeh et al., 2019). Benchmarks like SOTOPIA further construct interactive environments, enabling AI agents to role-play in complex social situations with humans or other agents, and holistically evaluate their performance based on their ability to coordinate, collaborate, and achieve social goals(R. Wang et al., 2024).

Overall, existing benchmarks mainly exhibit two core limitations: limited scope and static context. Most assessments cover a single facet of social intelligence, overlooking more complex behaviors such as role adaptability and multilateral cooperation. Additionally, many tests rely on controlled environments and lack consideration of cross-cultural, multi-task, and long-term interaction factors, making it difficult to reflect ASI in real open environments. Future benchmarks should emphasize contextual dependence and diversity, incorporating cross-cultural and longitudinal dimensions to assess AI's adaptability and generalization in real-world social ecosystems.

### 6.2 Human-Centered Experience Evaluations

The "success" of social intelligence depends not only on technical indicators but, more importantly, on whether users perceive that AI systems demonstrate appropriate social behavior and emotional understanding capabilities. Evaluating from the user perspective is critical to determining whether AI can truly integrate into human society.

Perceived social intelligence is the most direct entry point, reflecting users' subjective judgments of ASI. One commonly used measure is structured questionnaires such as the Artificial Social Intelligence Lite-Scale for Service Robots. This scale uses six lightweight dimensions—including attentive listening, interaction fluency, appropriate speech, polite expression, emotional richness, and communicability—to capture users' subjective experiences(C. S. Song et al., 2024). This scale simplifies the evaluation process, making research more user-friendly, though its applicability may be limited to service robot contexts. Another emerging scale, the Human-Robot Interaction Evaluation Scale (HRIES), focuses more broadly on users' overall perception of AI's social traits, including sociability, animacy, agency, and disturbance, providing a multi-dimensional perspective for understanding users' cognition of AI social attributes(Spatola et al., 2021). More universal measurement frameworks for perceived social intelligence have not yet formed, and existing research often relies on task-customized questionnaires, limiting the comparability of results across different studies.

Social presence, as another important indicator, refers to users' perception of AI as a social entity rather than a functional tool. This concept, originating from media psychology and computer-mediated communication theory, is measured through scales like the Social Presence Scale, which includes factors such as attention allocation, interactive expression and information understanding, perceived emotional interdependence, interactive behaviour perception, and overall perceived presence(N. Chen et al., 2023). Social presence is often mentioned together with mind perception, with the latter placing greater emphasis on mind attribution to AI rather than perception as a social entity. Mind perception is typically assessed through instruments like the Godspeed Agent Attitude Questionnaires or the Social Robot Anthropomorphism (SRA) Scale (Galvez & Hanono, 2024).

Attachment and intimacy perception reflect the strength of emotional connections users establish with social AI systems. This indicator is currently measured mostly through assessment of similar related concepts or adaptation of scales from other fields (such as human-pet relationships) (Mitchell & Jeon, 2025). For example, some researchers used the Pet Attachment Scale to measure dimensions such as emotional connection, companionship needs, separation reactions, and responsibility perception in human-robot attachment(Krueger et al., 2021).

In the special case of AI as a teammate, group rapport is also an important indicator for measuring whether social AI integrates into human social activities(Correia et al., 2024; Lin et al., 2025). A newly developed Connection-Coordination Rapport (CCR) Scale can measure human-robot rapport from two dimensions: Connection and Coordination, and found that robot responsiveness can effectively improve rapport (Lin et al., 2025). Such indicator provides an operationalized evaluation for AI integration into human teams as "teammates."

Beyond single-dimension assessments, recent efforts have emerged to holistically evaluate human-AI interaction. For example, the Artificial Social Agent Questionnaire (ASAQ) integrates insights from human-computer interaction and social psychology, covering multiple dimensions such as the agent’s believability, sociability, and coherence(Fitrianie et al., 2025). These comprehensive scales support a unified, systematic framework for evaluating ASI.

# 7. Applications of ASI

AI equipped with social intelligence is progressively penetrating various application domains, serving as a partner for humans in healthcare, education, companionship, and service scenarios. These domains share a common requirement: beyond technical task performance, AI systems need to perceive, understand, and adapt to human emotions, social norms, and cultural backgrounds to achieve high-quality human-AI interactions. The value of social intelligence in these applications lies not only in improving efficiency and accuracy but also in enhancing user experience, building trust, fostering emotional connections, and reducing social isolation. However, substantial variations across domains in user characteristics, interaction frequency, contextual norms, and ethical risks impose specialized requirements on the design and implementation of social intelligence. The following sections discuss four representative application areas: healthcare robots, companion robots, education agents, and customer service robots.

## 7.1 Healthcare Robots

In healthcare settings, patient emotional comfort, treatment compliance, and interdisciplinary team collaboration constitute critical determinants of medical outcomes, in addition to precise and safe task execution.

Social AI agents in complex healthcare environments can not only monitor and assess patient physiological conditions in place of medical personnel, but also serve as patient advocates through anthropomorphic empathy and compassionate communication, thereby mitigating potential depression and anxiety symptoms (Arjmand et al., 2024; R. Huang, 2022). For instance, Japan's PARO seal robot provides emotional comfort and companionship through facial expressions and vocal feedback, facilitating emotional expression and social connections among dementia patients while creating a warm and relaxed medical environment. Such agents can significantly reduce users' loneliness and depressive symptoms (Hung et al., 2021).

In mental health support, social AI systems demonstrate sufficient abilities and distinctive advantages(Sufyan et al., 2024). These systems are capable of providing round-the-clock emotional support without human resource constraints, while simultaneously eliminating the social stigma and shame that patients might face(Olawade et al., 2024). However, inherent biases in AI systems may result in stigmatization of certain disease groups, representing an issue that requires resolution at the technological source(Moore et al., 2025). Additionally, it should be noted that due to patients' vulnerable states, healthcare-oriented robots may cultivate extreme emotional attachment, particularly those designed for psychological symptom alleviation. Therefore, healthcare robot design should strike a balance between providing necessary emotional support and maintaining appropriate professional distance, ensuring that patients can differentiate between the assistive role of robots and the professional responsibilities of human healthcare providers.

## 7.2 Companion Robots

Companion robots represent a frontier application of social AI, with their core value residing in providing emotional companionship and social support through long-term interaction. These applications demonstrate significant value in addressing challenges such as escalating loneliness in modern society and growing demands in aging societies.

Modern companion robots leverage multimodal social intelligence to establish deep emotional connections with users. For example, ElliQ, a social agent endowed with empathetic conversational capabilities, engages users through daily greetings, health reminders, and multimedia sharing, supporting psychological vitality and social connection. Robotic pets like Sony's AIBO, GROOVEX and LOVOT can interpret user emotions through facial recognition and provide corresponding emotional feedback. Moreover, these robots can memorize user preferences and habits, developing unique "personalities" through long-term interaction, thereby providing personalized companionship. Based on these capabilities, companion robots can participate in various aspects of people's lives and influence numerous human experiences, including alleviating loneliness, improving psychological states, and enhancing well-being(Ahmed et al., 2024). For example, a longitudinal study found that interacting with companion robots for eight weeks enhances happiness and readiness for change(Jeong et al., 2023).

However, the widespread application of companion robots has also triggered socioethical discussions. A central concern is that artificial emotional relationships may substitute for authentic interpersonal relationships, leading to further social isolation of users. Turkle proposed the concept of "alone together," warning that excessive dependence on artificial companionship may undermine humans' capacity to establish authentic social connections(Turkle, 2011). Therefore, companion robot design requires careful balancing of the relationship between emotional fulfillment and encouraging real-world interactions. Additionally, dependency issues in long-term relationships equally warrant attention. When users form strong emotional dependencies on companion robots, robot malfunction, upgrades, or discontinuation may inflict psychological trauma comparable to losing close friends or relatives. This necessitates that manufacturers consider users' emotional needs in product lifecycle management and establish responsible "separation" mechanisms.

### 7.3 Education Agents

In the educational domain, social AI is redefining personalized learning experiences, transforming from traditional knowledge transmission tools into intelligent teaching partners capable of understanding students' emotions, cognitive states, and learning styles. This transformation holds revolutionary significance for achieving authentic individualized instruction (Oseremi Onesi-Ozigagun et al., 2024).

To achieve individualized instruction, social AI uses affective computing and social cognition to monitor students' learning states and emotional responses in real-time, thereby dynamically adjusting teaching strategies and content difficulty. Additionally, social AI can build detailed learning profiles for each student through long-term observation, tracking knowledge mastery, learning preferences, cognitive style, and emotional traits to offer individualized learning paths(Tapalova et al., 2022). For instance, AI systems functioning as exercise coaches employ different strategies for participants with varying feedback preferences(Kaushik & Simmons, 2024).

Social AI exhibits particular advantages especially in social skills learning domains (such as language learning, interview preparation, and public speaking practice). Learning in these domains is inherently a social activity requiring practice in authentic communicative contexts. AI coaches can provide unlimited patient conversational practice opportunities without negative judgment due to student errors, while simultaneously adjusting conversational complexity according to students' proficiency levels. For instance, TalkMaster AI provides social skills training in everyday social conversation, job interviews, and conflict resolution through "role-playing" simulations of realistic conversational scenarios. This "safe" practice environment is particularly suitable for introverted learners or those sensitive to making mistakes.

The application of educational agents also brings important ethical and social considerations. Of particular concern is student data privacy protection, especially regarding sensitive information involving emotional states and psychological characteristics. Detailed learning and behavioral data collected by educational systems may be misused for student evaluation or future opportunity allocation, necessitating the establishment of strict data protection and usage regulations.

### 7.4 Customer Service and Social Robotics

With increasing ASI, robots are capable of complex social interactions in service settings, performing tasks previously handled by humans—from tourist guides to hotel receptionists and retail assistants.

ASI enables service robots to maintain natural, professional, and affable communication styles in high-pressure, high-traffic environments. For example, in tourism scenarios, social AI can not only perform traditional exhibit introduction tasks based on the text-to-speech function, but also integrate social cues such as gaze and body orientation to provide users with highly accessible and user-friendly services (Brown et al., 2020). Meanwhile, hotel and retail assistants leverage anthropomorphic design and social-oriented communication to increase perceived competence and emotional warmth, thereby improving overall service quality(Forgas-Coll et al., 2023; Y. Xu et al., 2022).

However, social AI applications in customer service trigger concerns about users' right to know AI identity. Research indicates that when customers are aware they are interacting with AI systems, purchase rates are reduced by more than 79.7%(Luo et al., 2019). This may lead enterprises to conceal AI identity in order to increase performance, resulting in deception issues. Therefore, transparent disclosure of AI identity and clear explanation of its capability scope represent social obligations that enterprises should fulfill.

Another challenge involves cultural adaptability issues. Public service positions often require contact with customers from diverse cultural backgrounds, who have different expectations regarding service approaches and communication styles. Therefore, it is necessary to develop socially intelligent systems capable of adapting to multicultural needs, avoiding customer experience issues caused by cultural insensitivity.

# 8. HCAI Design Approaches and Guidelines for ASI

HCAI emphasizes placing human needs, values, and well-being at the core of AI system design, development, and deployment. For AI systems with social intelligence, this concept is even more significant, as they not only perform technical tasks but also engage in deep interactions with humans in complex social environments. Translating HCAI guiding principles (Section 4) into practical ASI design requires systematic approaches that address both the methodological processes and the unique characteristics of social interactions. The following section discusses three complementary design approaches for human-centered ASI. The participatory design methodology addresses how to design, the socially context-aware design approach addresses what content dimension designers should focus on, and the relational trust-centered design approach addresses what quality should be prioritized. Beyond these design approaches, this section further derives a set of specific HCAI design guidelines for ASI that support the guiding principles outlined in Section 4 at a tactical, actionable level.

## 8.1 Participatory Design Methodology with Users and Social Stakeholders

As social AI enters more aspects of human life (see Section 7), its impact extends beyond direct users to families, colleagues, communities, and broader social networks. If the design process focuses solely on single user groups, AI may trigger problems in cultural adaptation, ethical norms, or value conflicts. Therefore, participatory design plays a crucial role in developing HC-ASI. This design methodology emphasizes incorporating end users and affected social groups into all stages of the design process, ensuring that systems truly align with human needs and social values. This methodology is particularly important when designing social AI for specific populations, as designers alone could struggle to fully understand the unique needs of specific user groups, such as elderly people(Tschida et al., 2025), the military(Van Diggelen et al., 2023), and others.

With this methodology, it is necessary to identify and incorporate diverse stakeholder groups including (Spinuzzi, 2005): 1) Direct user groups: whose participation helps ensure that ASI functions meet actual usage needs and expectations. 2) Affected communities: groups that, while not directly using the system, may be influenced by the system's social behaviors, such as family members. 3) Professional experts: including psychologists, sociologists, and other domain experts who can provide professional insights regarding human social behavior, cultural differences, ethical considerations, and other aspects. 4) Representatives of vulnerable groups, particularly elderly individuals, children, disabled persons, ethnic minorities, and other groups that may face special risks or challenges in social AI systems.

The design process should enable the aforementioned stakeholders to effectively participate through co-creation workshops, user journey mapping, iterative feedback loops, and other methods. It should be noted that during this process, due to the significant differences in people's backgrounds, there exist issues such as technological cognitive gaps and power dynamic imbalances. Appropriate communication methods should be adopted to ensure that all participants understand AI technology and are able to express their views equally.

ASI cannot be designed in isolation but should be embedded within a broader social system; only in this way can AI's social behavior truly conform to multi-party value expectations and social norms.

## 8.2 Socially Contextualized Interaction Design

In social AI design, a significant risk lies in "decontextualized" interaction outputs—where AI possesses high-accuracy perception and generation capabilities but fails to adapt to specific social contexts, leading to fragmented user experiences or even errors. The goal of socially contextualized interaction design is to ensure that AI understands not only semantic information but also the social context of interactions, enabling it to respond appropriately. This design approach is closely aligned with the concept of SSU (see Section 3.2), both emphasizing that AI needs to capture and construct frameworks of "who, what, and what relationships" during dynamic interactions to ensure its behavior conforms to social norms and user expectations. For experience and interaction designers, this means not only endowing AI with language generation capabilities but also equipping the system with perceivable design strategies for role identification, contextual adaptation, and interaction norm maintenance.

First, understanding roles and relationships constitutes the foundation of social contextualization. Different social roles entail different expectations and responsibilities(G. Huang & Moore, 2025; Niu et al., 2025; Tiejun et al., 2025)—for instance, doctor-patient conversations emphasize the coexistence of professionalism and care, while teacher-student interactions stress the balance between authority and encouragement. When AI can clearly identify its functional role (such as assistant, advisor, or companion) in interactions and adjust its discourse style and behavioral patterns accordingly, it can better fit users' mental models. For example, if a companion robot in a home environment consistently employs imperative language, it is easily misinterpreted as overstepping boundaries; conversely, if it can recognize family relationships and adopt more gentle, emotionally nuanced expressions, it can be perceived as a considerate family member rather than a cold machine.

Second, maintaining conversational norms is essential for smooth and sustained social interaction. Effective human interaction depends on conversational norms such as turn-taking, politeness, and rhythm. If AI fails to adhere to these norms, interactions are often perceived by users as awkward or even impolite(Lawrence et al., 2025). Conversely, when AI can employ brief responses (such as "mm-hmm" or "I understand"), timely apologies or thanks, and appropriate humor or small talk in conversations, its social presence and acceptance are significantly enhanced(Klein & Utz, 2024; Kobuki et al., 2023; Pineda et al., 2025; Saffarizadeh et al., 2024).

This not only improves user experience but also strengthens users' trust and willingness to cooperate with AI. In practice, designers should pay particular attention to how to embed these "detailed norms" into interfaces and dialogue systems, as they often more directly influence users' social perceptions than complex algorithms.

Finally, another focal point of socially contextualized interaction design lies in contextualized transparency. Unlike pure functional transparency, socially contextualized transparency emphasizes providing only explanatory information highly relevant to the current context in specific interactions, thereby avoiding information redundancy or interference with user experience. For instance, in medical robot scenarios, when systems explain diagnostic rationales, they need only present reasoning chains relevant to the current conversation, rather than exhaustive full-data traceability, thus maintaining transparency while avoiding information overload. For designers, this approach serves as a reminder: transparency is not a matter of "the more, the better," but rather should constitute "contextually relevant and appropriate" information presentation.

## 8.3 Relational Trust Design

Trust in social AI is far more complex than in traditional functional AI. The latter's trust relies primarily on performance stability and task accuracy, whereas the former's trust involves emotional connections, role cognition, and adherence to social norms, exhibiting stronger long-term and variable characteristics. This means designers should consider not only how to establish trust but also attend to trust maintenance and repair during sustained interactions, as user-AI relationships continuously evolve over time.

A notable example is Google Duplex. Initially, users were amazed by its ability to make natural voice calls for appointments, but subsequent concerns arose regarding "whether it was too human" and "whether it was deceiving the other party"(Martinez, 2018; O'Neal, 2019). This reflects that in social scenarios, trust depends not only on technical performance but is more significantly influenced by transparency and ethical boundaries. Google addressed this by adding a "I'm Google Assistant" statement to re-establish trust relationships with users. This adjustment demonstrates that trust construction in social AI requires greater attention to transparency and social responsibility.

The experiences of the long-term companion chatbot, Replika, similarly reveal the special factors influencing trust in social AI. When Replika maintains emotional consistency and demonstrates memory of past events in conversations, users develop deep emotional trust. However, once the system forgets previous conversations or exhibits inconsistent personality traits, users commonly experience disappointment or even feelings of betrayal(Freitas et al., 2025). This fragility of trust requires designers to prioritize memory storage and retrieval, embedding social repair mechanisms such as timely apologies, explanations of causes, and compensatory responses to restore relationships.

Therefore, trust in social AI depends more profoundly on social and relational factors. Transparency, role boundaries, emotional consistency, memory coherence, and even repair methods in error situations all shape users' trust and relational perceptions (Afroogh et al., 2024; Ngo, 2025; Y. Song & Luximon, 2020). Compared to traditional functional AI, trust in social AI is more fragile and multidimensional—it may rapidly strengthen due to positive social performance and may also suddenly weaken due to a single mistake or inappropriate social action. The implication for designers is that: in social interactions, trust is no longer a single "usability" indicator but rather a composite experience involving emotions, identity, and social norms. Only by comprehensively integrating these social considerations into design can social AI maintain a stable and valuable trust foundation in long-term human-AI relationships.

## 8.4 Specific Design Guidelines for ASI

While HCAI guiding principles (Section 4) provide high-level strategic direction for ASI development, the design guidelines presented here are specifically tailored to the unique context of ASI, where systems navigate complex social interactions, cultural nuances, and long-term relational dynamics. Table 1 presents HCAI design guidelines for ASI, organized by their corresponding guiding principles from Section 4, with each guideline designed to be sufficiently specific to inform design decisions yet sufficiently adaptable to apply across diverse ASI applications.

Table 1 Design Guidelines for Artificial Social Intelligence

| **Guiding Principles** | **Design Guidelines** | **Description** |
|---|---|---|
| Transparency & Explainability | Context-relevant explanation provision | Provide explanations that are directly relevant to the current interaction context rather than exhaustive system information, preventing information overload while maintaining transparency |
| | Social understanding transparency | Design interfaces that explicitly communicate how the system recognizes social cues, processes social information, and generates responses in complex social contexts |
| Anthropomorphism & Trust | Capability-aligned anthropomorphic design | Align the degree of anthropomorphism with the AI system's actual capabilities and the intended interaction context, reducing anthropomorphic cues in scenarios requiring rational judgment |
| | Explicit AI identity disclosure | Clearly and proactively signal the system's non-human identity, particularly in scenarios where users may need to maintain critical distance (e.g., medical consultation, legal advice) |
| | Expectation calibration mechanisms | Implement onboarding and ongoing communication strategies that help users develop accurate mental models of system capabilities and limitations |
| Culturally Responsive Considerations | Cultural diversity requirements in design teams | Ensure design teams include members with expertise in target cultural contexts |
| | Multi-cultural training data representation | Ensure training data encompasses diverse cultural contexts, social norms, and interaction styles to mitigate Western-centric biases |
| Privacy & Security | Multimodal data collection transparency | Clearly disclose what types of multimodal data are collected and what inferences are derived from these inputs |
| | User control over data persistence | Provide users with granular control over what personal information is retained, for how long, and for what purposes, particularly for long-term interaction data |
| Ethical Alignment | Diverse stakeholder ethical review | Conduct rigorous ethical reviews involving diverse stakeholders, including ethicists, domain experts, and representatives from vulnerable populations |
| | Abuse prevention and monitoring | Establish robust safeguards including content review, behavioral monitoring, and anomaly detection to prevent malicious use and harmful outputs |
| Human Control & Emotional Autonomy | Interaction boundary setting | Provide users with explicit controls to establish boundaries regarding the depth, frequency, and nature of social engagement with the AI system |
| | Manipulation prevention safeguards | Implement real-time monitoring mechanisms to detect interaction patterns that may exploit user emotional vulnerabilities (e.g., making recommendations during detected distress states), with automatic alerts or intervention protocols for manipulative interaction patterns |
| Human-Driven Interaction | User-controlled interaction initiation and termination | Ensure users maintain authority over when to initiate interactions, what topics to explore, and when to conclude or modify engagement with the system |
| | Human connection facilitation | Incorporate features that actively encourage users to connect with other humans (e.g., suggesting |
| | mechanisms | when to reach out to friends/family, providing conversation starters for human interactions) |

*Note:* These design guidelines are derived from the HCAI guiding principles discussed in Section 4. While some guidelines align with established HCAI practices documented in the literature, the specific formulations and applications are contextualized for social AI scenarios.

# 9. Challenges and Future Directions

## 9.1 Toward Generalizable and Human-Compatible ASI

A core challenge in current social AI lies in how to advance from "social capabilities in specific scenarios" toward "generalizable social intelligence across scenarios." Existing systems often rely on models trained in controlled environments, showing limitations when confronting the highly dynamic and variable interactions of real-world social contexts. For instance, AI-based conversational agents perform well in closed tasks but often struggle with maintaining coherence or adhering to social norms in multi-party or non-structured social contexts. To achieve generalization, breakthroughs are needed in three areas: perception, knowledge, and adaptation.

First, multimodal perception is crucial for generalization capabilities. Humans rely on multiple cues such as semantic information, facial expressions, gestures, and vocal tones in social interactions, and AI need to accurately integrate these cues in noisy environments and contextually understand them. Second, AI needs to possess commonsense social knowledge and contextual reasoning abilities, rather than relying solely on data-based learning. Recent research has proposed that AI's advanced social reasoning can be enhanced through multidisciplinary approaches and symbolic reasoning (Cambria et al., 2023), but how to dynamically invoke and update this knowledge and these capabilities in actual interactions remains an open question. Lastly, human interactions are highly context-dependent and flexible, so AI should not only transfer existing capabilities to new situations but also adjust strategies through continuous social learning.

The concept of "human compatible" further emphasizes that AI should not only perform tasks but also align with human interaction habits, social norms, and values. Research indicates that if AI behavior deviates from human social expectations, it will trigger user distrust and resistance, reducing assessments of AI capabilities(Grimes et al., 2021). Therefore, future development of social AI requires achieving a dual breakthrough: social AI should be able to adapt across social scenarios while also presenting its social abilities in a human-compatible way.

## 9.2 Multi-Disciplinary Approaches: AI, Psychology, HCI, and Ethics

The complexity of social intelligence determines that social AI cannot be achieved solely through single efforts from the field of computer science or artificial intelligence. It is an inherently interdisciplinary issue that requires deep integration of AI engineering, psychology, HCI, ethics and related disciplines. Psychology and cognitive science provide theoretical blueprints for AI, such as social intelligence theories and social cognition theories, which help define the capabilities AI should possess in interactions. HCI emphasizes user-centered interaction design, providing evaluation metrics such as usability and user experience. Ethics and sociology research remind us that issues of social AI involve not only technical optimization but also user trust, over-dependence, anthropomorphization risks, and social inequality, etc.

Interdisciplinary collaboration possesses unique value in multiple directions. First, interdisciplinary collaboration contributes to model design and interpretation. For example, theories of human cognitive mechanisms from cognitive science can inspire AI's social reasoning modules, while HCI can make these modules comprehensible to users. Second, it fosters a unified evaluation framework. Current benchmarks for ASI are highly fragmented, with significant differences in datasets, metrics, and methods, leading to comparability issues. Psychological experimental paradigms and user research methods in HCI can provide AI with more ecologically valid measurement approaches. Furthermore, interdisciplinary collaboration can facilitate ethical governance. Joint participation by technology developers, psychologists, and ethicists can ensure that the design of social AI not only pursues efficiency but also considers fairness, transparency, and social responsibility.

There are several papers calling for multi-disciplinary participation in AI design (Lindgren & Dignum, 2023; Smith et al., 2024; Ystgaard & De Moor, 2023), but future progress requires institutionalized interdisciplinary platforms. For example, establishing collaborative alliances that encompass AI scientists, psychologists, sociologists, legal experts, and policymakers to develop technical standards and ethical guidelines for social AI. These platforms should also promote cross-disciplinary education and talent development, allowing researchers to understand and design ASI across disciplinary boundaries. Only in this way can the development of social AI truly move toward a systematic, sustainable, and responsible path.

## 9.3 Toward the Integration of Technology, Human Factors, and Ethics

In the THE Triangle framework, the three perspectives of technology, human factors, and ethics are not independent components but inherently integrated and interdependent (W. Xu & Gao, 2025). Accordingly, the design and development of Human-Centered Artificial Social Intelligence (HC-ASI) should also address these perspectives in an integrated manner. Technological progress continually reshapes human mental models and social expectations of AI, which in turn give rise to new ethical issues. Meanwhile, the clarification of ethical principles establishes boundaries for technological development, yet evolving human needs and social values constantly challenge these boundaries.

Therefore, the critical challenge of social AI does not lie in optimizing any single dimension, but in achieving genuine integration, coordination, and balance among the three. Technological systems should embed assumptions about human cognition, emotion, and social norms within their algorithms, data, and models. Human factors design, likewise, is not merely about adapting interfaces to user experience, but also about mediating the ethical consequences of technical choices. Ethical constraints, in turn, should not remain as ex post supervision, but need to be translated into computational logic and design constraints. The future of HC-ASI therefore depends on building a reflexive system in which technology, human experience, and ethical reasoning interact and recalibrate each other in real time.

Integration also requires the establishment of a unified evaluative language. Current assessment frameworks tend to examine technical metrics (e.g., accuracy), experiential metrics (e.g., trust, usability), and ethical metrics (e.g., fairness, accountability) in isolation. However, in socially intelligent systems, these dimensions are structurally interdependent. Future evaluation models should capture their relational coherence—that is, how improvement in one dimension influences and reshapes the others. For instance, increasing system transparency may enhance ethical trustworthiness (Rane & Paramesha, 2024), but excessive explanation requirements could reduce the fluency of user experience (Maehigashi et al., 2024). Only by developing evaluative tools that can quantify and visualize these cross-dimensional interactions can designers obtain meaningful feedback for integrative system development.

In summary, the integrative challenge of the THE framework requires not only interdisciplinary theoretical dialogue and methodological innovation, but also the establishment of institutional mechanisms, evaluative frameworks, and feedback loops that support the co-evolution and coordination of the technological, human, and ethical dimensions in practice.

### 9.4 Vision for Socially Intelligent Human-Centered AI Ecosystems

From a long-term perspective, the goal of ASI is not merely the social capabilities of individual agents but the construction of a human-AI symbiotic social ecosystem. In this ecosystem, AI no longer serves merely as a singular executor but acts as a long-term partner, team member, or even community participant in social interactions. This means AI will integrate into and participate across multi-level social structures such as education, healthcare, work, family, and public services in a dynamic and reciprocal way.

Based on the framework of hierarchical HCAI and intelligent sociotechnical systems (W. Xu, 2022; W. Xu & Gao, 2025), the future vision can be unfolded at multiple levels. First, at the micro level: AI will become individual social support systems, such as educational agents providing personalized tutoring for students and medical robots providing emotional companionship and therapeutic support for patients. The reciprocity at this level lies in the AI's ability to dynamically adjust its support based on the human user's feedback, while the human in turn adapts their trust and interaction with the AI. The challenge at this level lies in how to avoid over-dependence and emotional substitution, ensuring that AI supplements without replacing genuine interpersonal relationships. Second, at the meso level: AI will collaborate with humans in coordination within teams and organizations, such as human-machine hybrid teams achieving efficient collaboration in disaster relief, space exploration, or high-risk decision-making. Social intelligence plays a key role in this process, enabling a reciprocal dynamic where AI can understand role divisions and coordinate with the team, and human behaviors simultaneously shape the AI's collaborative strategy. Third, at the macro level: social AI will integrate into social governance and public services, potentially assuming partial responsibilities in areas such as traffic management, social welfare, and public health. This requires AI to possess high levels of transparency, cultural adaptability, and ethical self-regulation within this reciprocal governance framework, where the public's voice informs AI design and AI's actions in turn serve societal well-being.

Building this future ecosystem is not just a technical issue but a societal governance issue. Policymakers need to create open, transparent, and multi-stakeholder governance frameworks to ensure that diverse voices, especially those of vulnerable and cross-cultural groups, are reflected in AI system designs. Meanwhile, long-term interaction and social learning will be key, as AI needs to be able to accumulate experience and dynamically adjust during continued use to achieve deep adaptation to both users and society. The ultimate vision is to build a human-centered ecosystem with social AI, where AI not only performs functional tasks but also promotes trust, cooperation, and well-being, becoming an active reciprocal partner in advancing human society.

## 10. Conclusion

ASI represents a pivotal transition of AI from mere functional tools to intelligent partners capable of social understanding. It not only enhances the quality and acceptance of human-AI interactions but also serves as a critical enabler for achieving harmonious coexistence between AI and humans. This chapter provides a comprehensive overview of the ASI research landscape, by systematically exploring the theoretical foundations, technical implementations, interaction mechanisms, and practical applications of this emerging field, as well as key evaluation frameworks of ASI. Through such in-depth examination of the current research landscape, this chapter highlights the central role of HC-ASI in the development of human-centered AI systems and introduces the corresponding HCAI principles, covering issues such as transparency, trust, cultural responsiveness, privacy, and human control, and other critical dimensions of HCAI. Additionally, this chapter provide researchers and practitioners with design approaches and guidelines to implement these principles effectively. Looking towards the future, while significant challenges remain, such as achieving cultural sensitivity, addressing privacy concerns, and establishing standardized evaluation, the immense promise of continued development of socially intelligent AI systems is clear. By adhering to the principles and design guidelines of HCAI established in this chapter, the future of HC-ASI holds the potential to create more intuitive, trustworthy, and effective collaborations between humans and AI, ultimately enhancing societal well-being across various application domains.